\documentstyle{article}
\newpage
\begin{document}

\begin {center}
{\Large \bf  {Remote Observing and Experience at ESO}}
\end {center}
\vskip0.5truecm
\begin {center}
A. Wallander and A.A. Zijlstra
\end {center}
\vskip0.5truecm
\begin {center}
European Southern Observatory, Karl Schwarzschild Strasse
2,\\
D-85748 Garching bei M\"unchen, Germany\\
\vskip0.1truecm
email: {\tt azijlstr@eso.org}
\end {center}
\vskip2.0truecm
\begin{abstract}
Remote observing can be broadly defined as those observations where
the astronomer is not physically present at the telescope.
Different implementations presently in use include robotic
telescopes, service observing with or without eavesdropping and
active remote observing. We briefly describe the terminology, the
pros and cons, the observing modes, and their implementation at
optical observatories.

In the second part of the paper, we discuss the example of remote
observing with ESO's NTT.  Different aspects of the technical setup
and the support given to observers, with emphasis on problems
encountered, are described.  With the present system, we find that
the observing efficiencies for local and remote observing are
identical: few projects still require local observations.

\end{abstract}

\section{History}

Remote observing has been widely used in radio astronomy for the
last decade. However, in the optical and infrared domain very few
observatories have been successful in supporting it.  One reason for
this difference is that optical/infrared telescopes are more
demanding in the sense of object acquisition and amount of science
data produced.  Another reason could be that optical astronomers
have less trust in the performance of the telescope and instruments.

In the early 1980's a number of observatories started experimenting
in remote observing (Raffi \& Tarenghi 1984, Longair et al. 1986,
Raffi \& Ziebell 1986).  Most of these early attempts failed,
mainly because the technology was not yet matured.  The
communication links were not reliable and the data transmission
rates were too low. It also became evident that it was very
difficult to operate a telescope originally designed for local
control in remote mode.

In the late eighties the situation became more favourable. First,
there was a general improvement in communication infrastructure:
cheaper, more reliable and faster, not to forget the development of
Internet and, more recently, of the World-Wide Web. Second, a new
generation of 3--4-m class telescopes was designed and went into
operation. For some of these, remote observing had been foreseen
already during the design phase (Loewenstein \& York 1986, Raffi et
al. 1990). This led to the successful use of several different modes
of remote observing, ranging from  fully automatic telescopes to
interactive long-distance instrument and/or telescope control.

In 1992 a workshop dedicated to remote observing was held in Tucson
(ed. Emerson \& Clowes 1993), where many of the available forms were
extensively discussed.  We will first summarize the terminology
adopted at this conference for the various remote-observing modes,
before discussing the pros and cons and the technical requirements
for remote observing. In the second part of the paper we discuss the
experience obtained at ESO, with emphasis on the observing
efficiency, the support structure and the feedback from astronomers.

\section{Definitions of Remote-observing Modes}

\hskip \the\parindent {\it Robotic Telescopes} - A telescope and
instrument which is programmed beforehand and needs little or no
interaction during the night. An entire night's observing program
may be downloaded during the day and the scientific data uploaded
during the observations or later.

{\it Remote Engineering} - A remote engineer performs
engineering activities, e.g. installations, diagnostics,
troubleshooting, on local equipments.  Although this is not an
observing mode it should be included in the context, because if
remote engineering is available, remote observing may come for free.
Remote engineering can be very effective, especially since often
engineers need to come to the telescope for relatively minor tasks
and the travel distance can be significant.

{\it Service Observing} - An astronomer fully specifies the
objects to be observed and the instrument configuration beforehand,
and the actual observation is carried out by the observatory staff.
The astronomer is awarded observational data rather than telescope
time.

{\it Passive Remote Observing} - A remote observer monitors an
observation carried out at a telescope by a collaborator or service
observer.  The remote observer has access to the data obtained and,
optionally, observation parameters, and interacts with the local
observer via voice, ``talk'', or e-mail. This is also called {\it
Eavesdropping}. It is probably fairly common, but observatories tend
not to keep track of this.

{\it Active Remote Observing} - A remote observer interactively
controls an instrument and, optionally, a telescope at an
observatory. In most cases a local telescope operator is required
for safety reasons and sometimes to operate some telescope and
auxiliary equipment. This is also called {\it Remote Control}.

Combinations of these observing modes are of course possible, and
may in fact give additional advantages.

{\it Robotic telescopes} are by far the most popular of these
options. They are typically small telescopes (50cm) and there are
two important uses: the first is for long-term monitoring programs
(an example is the network of small telescopes looking for solar
oscillations) and the second is for educational purposes: some
universities now have a small robotic telescope where astronomy
students can send in requests electronically for short observations.

At the other extreme, {\it active remote observing} is rare. The
only large telescope for which this mode is the norm is the 3.5m on
Apache Point Observatory (APO), which is run by a consortium of US
universities. This telescope is operated for 80\% of the time
available for scientific observations in remote control. Typically,
a night will be split between several projects, and each astronomer
can carry out the observations from his or her campus.  ESO
has two telescopes which can be operated in remote control: the 1.4m
CAT which is used for high-resolution spectroscopy mainly of bright
stars, and the 3.5m NTT.

\section{Pros and Cons}

The motivation for remote observing has been debated for some time
in the literature and a high level of agreement on the main
arguments has been obtained (e.g. Emerson \& Clowes 1993). These are
summarized in Table 1 and briefly discussed below.

\begin{table}
\caption{Weighing the pros and cons of remote observing}
\begin{tabular}{||l|l||l|l||}   \hline \hline
Proven         & Likely         & Likely          &  Proven
 \\
advantages     & advantages     & disadvantages   &  disadvantages
 \\
\hline \hline
Flexibility    & More than one  & More difficult  &  Expensive
  \\
               & participating  & to concentrate  &  - personnel
  \\
               & astronomer     & on observing    &  - communication
 \\ \hline
Shorter obser- & May save costs & Not as efficient&
 \\
ving programs  & in some cases  & as classical    &
 \\ \hline
Convenient for &                &                 &
 \\
astronomer     &                &                 &
 \\
\hline \hline
\end{tabular}
\end{table}

The main arguments in favour of remote observing are all related to
the fact that the astronomer does not need to travel to the
observatory. This is a strong argument if the observatory is in an
inaccessible place, the most extreme example being the planned
observatory on the Antarctic Plateau for which remote observing is
envisaged (Burton 1995). It also allows for the observing schedule
to be made at short notice (as is done at the VLA). However,
flexible and queue scheduling is not recommended in active remote
control and/or eavesdropping: for the astronomer to be directly
involved the time of the observations needs to be known in advance.
Instead it is possible to schedule shorter observing programs, in
units of hours instead of nights.  Monitoring
programs also become easier to schedule without an astronomer at the
observatory having to be involved. The elimination of long travel
times and acclimatization results in savings of astronomer's time --
an important point for researchers at universities who may have
teaching duties.

Remote observing very often allows more than one astronomer to take
part in the observations. The impression at ESO is that, for the
large telescopes, astronomers prefer to come with more than one
person: while one person does the observing, the other concentrates
on the on-line data analysis.  For remote control, financial
support for an extra observer is easier to arrange. ESO also
attracts some Eastern-European astronomers for remote observing
who could not possibly afford to travel to Chile. For them remote
observing can save costs, but for the observatory this will in
general not be true unless the required communication links are
free, or very cheap, and little additional support is required.

Cost is the main argument against remote observing. In order not to
degrade the scientific efficiency, the remote observer must get the
same support as the local observer. This means that often the
support personnel have to be duplicated. Sufficient bandwidth is
also required in order not to have idle telescope time and the
cost of this bandwidth can very easily overtake the savings in
travel cost.  Table 2 shows typical cost estimates for the case of
ESO, illustrating that remote observing only reduces total costs
when it is used for a significant fraction of the observing
proposals for at least two telescopes.  A second argument against
remote observing is, that, due to limitations in bandwidth, there
is a time delay before the observer sees the data coming from the
instrument. (This problem is not unique
to remote observing but is experienced by everyone observing with
CCDs.)  This becomes a problem whenever the extra delay is a
significant fraction of the average integration time per exposure.
Finally, if the observations are done from one's own office, the
distractions due to the normal office activity may easily cause a
loss of efficiency.

\begin{table}
\caption{Comparing operational cost (in kDM/month)}
\begin{tabular}{||l||l||l|l|l|l||}   \hline \hline
No. of telescopes & 0\% RO  & 10\% RO & 30\% RO & 50\% RO &
 100\%RO\\
and observing runs& 0 rem op& 1 rem op& 2 rem op& 2 rem op&
 3 rem op\\
\hline
1 tel.5 runs/mon. & 20      & 46      & 53      & 50      & 52
     \\
1 tel.10 runs/m.  & 40      & 61      & 68      & 62      & 56
     \\
1 tel.20 runs/m.  & 80      & 102     & 100     & 86      & 64
     \\
CAT+NTT 20r/m     & 80      & 102     & 100     & 86      & 64
     \\
\hline
                  & 0 rem op& 1 rem op& 3 rem op& 4 rem op&
 6 rem op\\
\hline
4 tel.80 runs/m.  & 320     & 322     & 291     & 250     & 142
  \\
\hline \hline
\end{tabular}
{\footnotesize
The following assumptions are used for the calculations:\\
Classical observing run (travel, accommodation etc.) = kDM 4\\
Remote observing run (travel, accommodation etc.) = kDM 0.8 \\
Remote operator (rem op) = kDM 10/month\\
Link cost (50\% of total cost) = kDM 18/month\\
Same local support for classical and remote observing}
\end{table}

\section{Requirements and Techniques}

The main requirement for all modes of remote observing except
robotic telescopes is a fast data transfer from the observatory to
the remote site. Data files produced by modern instruments are
large (a typical CCD frame is 2 to 8 Mbytes), and they should
ideally be transferred in no more than a few minutes.  To achieve
this a fairly high bandwidth is required, although modern data
compression algorithms can improve
the transfer rate.  Compression based on the H-wavelet transform has
been successfully applied at a number of observatories.  Both the
availability and the reliability of the communication link need to
be very high.

A second requirement is that the telescope and instrument must be
reliable and stable. Solving technical problems is generally more
time consuming during remote observing, due to the extra feedback
time.  A good way to limit unforeseen problems is by keeping
instrument change-overs to a minimum. Limiting the instrumentation
is also a better way to ensure adequate know-how at the
remote-observing site.

Internet is certainly the main carrier of remote observing traffic.
The tremendous advances in bandwidth and number of users and, more
important, the fact that Internet is still ``free'', makes it the
obvious first candidate to implement the communication link.  The
World-Wide Web gives a convenient interface which is already used
for some robotic telescopes. A future commercialization of Internet
may make some of these advantages obsolete. Also, for observatories
located in remote places it may be difficult to obtain a fast
Internet access. As an alternative, dedicated links to more
populated areas with Internet access can be acquired.  Although
dedicated links are more expensive they also offer advantages like
guaranteed bandwidth and propagation delays.

A {\it Remote Observing Center} is a dedicated geographical site
from where the observation is carried out. This is typically
located in a major astronomical research institute and the remote
observer has to travel to this site.  The advantages are that
expert knowledge, similar to a local observatory, can be built up
in order to provide accurate support to the visiting astronomer.
{\it Distributed Remote Observing} is when any site with a network
connection, normally Internet, can perform remote observing. The
obvious advantage is that no travel is required. However, it
assumes that the remote observer has expert knowledge, and does not
need dedicated personal support.

\section{ESO Experience}

ESO currently supports {\it active remote observing} with two
telescopes.  The 1.4m CAT has been operated routinely in
active-remote-observing mode about 50\% of the time for the last six
years (Baade 1993).
The NTT was the first ESO telescope specifically designed to
accommodate remote observing. Remote observing was first offered to
the user community in 1993, and since then about 15\% (four to five
nights per month) of the time is scheduled in
active-remote-observing
mode. Early experiences with the system are described in Balestra et
al. (1992), Baade at al. (1993) and Wallander (1994).

At ESO Headquarters in Garching a {\it Remote Observing Center}
provides the interface for remote observers to the La Silla
Observatory in Chile.  Observing from other sites and institutes is
not supported.  A dedicated satellite link provides the bandwidth
between the two sites.  The system provides fully interactive {\it
active remote observing} based on dedicated remote software
executing at the remote site. The architecture of the NTT remote
observing system was reported in detail in Wallander (1993).

In November, 1994 the $64\,$kbps (used by NTT) and the analogue
(used by CAT) PTT-leased communication links were replaced with a
$2\,$Mbps ``roof-to-roof'' satellite link.
The complete system was obtained as a turn-key project
from an external contractor.  Although this involved a considerable
capital investment, the actual operation costs did not increase. It
turned out that by leasing the bandwidth directly from the satellite
provider, instead of via national PTT's, a 26-fold increase in
bandwidth could be acquired for a lower price.  In addition, by
mounting the antennae directly on the ESO premises and becoming
independent of PTT's, the reliability of the link increased.  During
the period November 1994 to June 1995 the availability of the link
has been over 99.8\%. This should be compared to the record of the
previous $64\,$kbps PTT-leased line, which in the period 1991 to
1994 had an average availability of 95\%.

Figure 1 shows the data transfer rate from La Silla to Garching,
averaged every five minutes, over the last ten months. The
saturation of the previous $64\,$kbps link, of which 32--48$\,$kbps
was allocated for data, is clearly visible. The zoom of the first
two weeks of April shows that remote observing is a main, but not
the only, user of the link.

\begin{figure}
\vspace{10cm}
\caption{ Link utilization }
\end{figure}

\section{Observing Efficiency}

We define the observing efficiency as the fraction of time between
nautical twilights that the shutter of the instrument is open.  This
number is automatically derived from the computer-based operation
log and we have routinely recorded it for the NTT since November
last year.  We only used data from nights without any technical
problems or time lost for weather, and where the instrument, EMMI
or SUSI, was used in one of the standard modes. The result is shown
in Figure 2.

With the exception of one remote observing run (2 nights) the
obtained efficiency for classical and remote observing are very
similar. Performing some statistics on the data we get the result
shown in Table 3, where the first number gives the efficiency
expressed as a percentage, with its standard deviation, and the
number in braces gives the number of nights used in the
calculation. The first night of a run is listed separately because
of possible familiarization effects (``first-night syndrome''),
which are indeed present.
\begin{figure}
\vspace{10cm}
\caption{Comparing classical and remote observing efficiency }
\end{figure}

\begin{table}
\caption{Comparing classical and remote observing efficiency}
\begin{tabular} {||l|l||l|l||} \hline \hline
 \multicolumn{2}{|c|}{Classical observing} &
 \multicolumn{2}{|c|}{Remote observing}  \\
  First night &   Later nights  & First night   &   Later nights
 \\
\hline \hline
64+/-11\% (22)&   71+/-10\% (29)& 60+/-18\% (11)&   67+/-18\% (8)
 \\
              &                 & 65+/-10\% (10)&   73+/-9\% (7)
 \\
\hline \hline
\end{tabular}
\end{table}

Due to the still small number of nights, the difference between
remote and local observing (a few per cent) can entirely be
attributed to the one observing run which reached very low
efficiency (which explains the larger standard deviation).
Removing this run we get identical figures. We conclude from this
that for normal observations remote observing is now competitive.

\section{Observing Support}

The support given to the remote observer requires special attention.
Socially, there are more distractions making it more difficult to
concentrate on observing. The time difference between La Silla and
Garching is such that the second half of the night in Chile
corresponds to the morning in Garching, when office life goes on as
usual.  Having a separate observing room is a necessity.  It is also
more difficult to locate people with adequate technical knowledge of
the system who can introduce the observer and help with preparing
the observations. This is especially true if the remote observing
system is only used occasionally.

It should be noted that very few people have so much observing
experience on the NTT that they do not need any introduction. In
practice, most astronomers only use a particular telescope a few
nights per year, and they will probably try to use a different mode
from the one before. In order to allow the observer efficient use of
the telescope, it is essential to arrange for good support. For the
NTT, we now have a group of four post-doctoral fellows and students
who give introductions in Garching.  During the observing night the
astronomer is also supported by a remote-control operator in
Garching, who controls the telescope, and a local night assistant
present at the telescope.  The latter is present only as a
safeguard, but essential to assist in case of problems.

One could describe the support problem as trying to build an
observatory-environment away from the telescopes. People need to
have the know-how and be aware of recent problems and changes (which
requires good communication channels with the observatory). At ESO,
the introduction of a ``telescope team'' for the NTT, with staff
both from Garching and La Silla, has helped significantly in
improving communications. The effort put into the support
structure is considerable, but is necessary if remote observing is
to be more than a tool for highly experienced observers, which most
astronomers are not.

The experience at ESO is that in almost all cases, the home
institute pays for a second observer in addition to the one paid
for by ESO, in the case of remote control. This 'self-support' also
helps to improve the efficiency of the observations. It may be of
importance from an educational point of view as well.

\section{Perceptions by Astronomers}

Astronomers who have used the remote control system are often, but
not always, satisfied. Often their feedback has resulted in
improvements of the system or of the support.  When applying for
observing time with the NTT, the astronomer has to specify why the
program would not be suited for remote observing. We do not force
people to use remote observing against their will, which would
certainly backfire, but the remarks give a good impression of the
perceived problems with remote observing.

Occasionally, the overheads are thought to be larger in remote
observing. This may be true for complicated programs, and is
certainly true if the observer waits for the image to arrive before
deciding what to do next, as some do. Part of this problem has been
solved with the new, much faster link.

More commonly, the astronomer feels that the judgment whether the
conditions are photometric are more difficult in remote observing
mode.  It is not so clear whether this argument is correct. It is
always difficult to judge how photometric conditions are when using
CCDs, especially during dark time when no moon is present. The
La Silla night assistant keeps an eye on the weather, but
astronomers have been reluctant to accept his opinion. There have
been cases where the observer wanted to continue while the moon was
no longer visible. Data from the meteomonitor, which displays
seeing, windspeed, temperature, etc., are displayed on-line and
this is one of the most popular things to watch in the control room.

Twilight flats are seen as the most demanding part of the
observations, because there is little time available and the
observer cannot wait for the image to arrive before knowing what
the count levels were in the previous flat. The solution here is to
do fast
statistics on the data on the La Silla workstation, where the
original file first arrives.  This is perfectly possible, either by
the observer working over the link or by the night assistant on La
Silla who sees each image immediately after it has been read out.
Good results have been obtained by using the Tyson sequences for
twilight flats (Tyson \& Gal 1993).  The program to calculate these
sequences is available on-line.

Some astronomers also express doubt about focussing. The instrument
is generally focussed using one or more exposures of a random field
(preferably close to the target). Here the image transfer can be
very quick because it is not normally necessary to read out the full
CCD. The image is normally analysed by the night assistant in La
Silla, but it has been done remotely as well. This took 30 seconds
longer and gave the same answer.

The most serious argument used by astronomers is that their object
is very faint and that it is difficult to position it at the slit.
If the object is not visible on the video screen, it is necessary
to take acquisition images and to move the telescope such that the
selected object falls in the slit. Often two such exposures are
needed. The additional overhead becomes larger if the science
exposures are relatively short and there are many targets: if many
acquisition images need to be taken the observer should go to La
Silla.

\section{Future}

What is the future of remote observing? There is clearly a demand
from the ESO community, seeing that a third of the optical NTT
proposals request this mode.  Part of this may be due to the fact
that it is easier to come with more than one person in this mode.
The recent experience has shown that, for many programs, remote
control is competitive with local observing, being as efficient in
telescope usage while giving a saving of the astronomer's time.  At
the same time, there is a large group of people who prefer to
travel to La Silla. We will for the time being continue to offer
remote observing as a service to the community, but not force it
upon people.  We will try to improve the system to alleviate the
doubts as expressed above.

A major upgrade of the NTT control system is being undertaken as
part of the NTT Upgrade Project. The aim of this activity is on the
one hand to verify the concept and software to be used for the VLT,
on the other hand to provide an identical interface on the NTT for
higher level operational tools, procedures and methods to be used
on VLT. It is expected that the VLT technology and software
architecture will give essential performance advantages also for
remote observing. Faster computers, more efficient communication
protocols, on-the-fly data compression and fast data forwarding
will reduce the data-transfer rate.  The limiting factor of a CCD
display will become the readout time, independent of where the
display unit is located.

\section{Conclusions}

We have shown that the observing efficiency does not degrade when
using {\it active remote observing} for the ESO NTT as compared to
classical observing.  This allows more flexibility in scheduling,
shorter observing programs, long term monitoring programs, and
savings of astronomer's time.

However, {\it active remote observing} is nothing else than moving
classical observing to another site.  It does not address the
``first-night syndrome''.  To increase the scientific efficiency,
service observing may be a more important observing mode than remote
observing. Assuming the service observer will be at the telescope,
we would expect increased demands for {\it eavesdropping}
capabilities. The requirements for this to be successful are a
sufficiently fast link and adequate communication facilities, i.e.
not much different from those of {\it active remote observing}.
The main role of {\it active remote observing} may be found in the
new generation of large telescopes, where the observing runs may be
very short, and for astronomers in places where travel money is
difficult to get.

\vskip 20pt
\noindent
{\bf Acknowledgements} Manfred Ziebell, with support from Joar
Brynnel, has been responsible for the very successful installation
and operation of the new satellite link.  We thank Miguel Albrecht
for providing the observing efficiency data and Jesus Rodriguez for
helping us obtaining information from several observatories.  The
successful operation of ESO remote observing would not have been
possible without dedicated support from the whole operation crew,
both at La Silla and in Garching.

\section{References}
\begin{list}
{}{\itemsep 0pt \parsep 0pt \leftmargin 3em \itemindent -3em}

\item Baade D. 1993, Observing at a Distance, ed.  Emerson
D., Clowes R. (World Scientific, Singapore), p. 131
\item Baade D., et al. 1993, The ESO Messenger, 72, 13
\item Balestra A., et al. 1992, The ESO Messenger, 69, 1
\item Burton M. 1995, in ASP Conf. Series, 73, p559-562
\item Emerson D., Clowes R. 1993, Observing at a Distance,
       (World Scientific, Singapore)
\item Loewenstein R.F., York D.G. 1986, SPIE 627
\item Longair M.S., Stewart J.M., Williams P.M. 1986,
       Q. Jl R. Astr. Soc. 27,153
\item Raffi G., Tarenghi M. 1984, The ESO Messenger 37, 1
\item Raffi G., Ziebell M. 1986, The ESO Messenger 44, 26
\item Raffi G. et al. 1990, SPIE 1235
\item Tyson N.D., Gal R.R. 1993, AJ, 105, 1206
\item Wallander A. 1993, Observing at a Distance, ed.  Emerson
D., Clowes R.  (World Scientific, Singapore), p. 199
\item Wallander A. 1994, Nuclear Instr. and Methods in
       Physics Research Vol. A347, 258
\end{list}

\end{document}